\documentclass[11pt]{article}
\usepackage{fullpage}
\usepackage{amsmath}
\usepackage{amssymb}
\usepackage[dvips]{epsfig}

\def\##1{\underline{#1}}
\def\=#1{\underline{\underline{#1}}}

\def\+#1{\underline{\bf #1}}
\def\*#1{\underline{\underline{\bf #1}}}

\def\r#1{(\ref{#1})}
\def\l#1{\label{#1}}
\def\c#1{\cite{#1}}

\def\le{\left(}
\def\ri{\right)}
\def\les{\left[}
\def\ris{\right]}

\def\ric{\right\}}

\def\.{\mbox{ \tiny{$^\bullet$} }}

\def\epso{\epsilon_{\scriptscriptstyle 0}}

\def\muo{\mu_{\scriptscriptstyle 0}}

\def\curl{\nabla\times}

\def\curl{\nabla\times}

\def\tr{(ct,\#r)}
\def\ok{(\omega/c,\#k)}

\def\mrbh{m_{\mbox{\tiny{bh}}}}
\def\arbh{a_{\mbox{\tiny{bh}}}}
\def\qrbh{q_{\mbox{\tiny{bh}}}}

\def\Dkn{\Delta_{KN}}
\def\dkn{\delta_{KN}}
\def\Dks{\Delta_{KS}}
\def\dks{\delta_{KS}}

\def\St{\tilde S}

\begin{document}

\begin{center}

{\bf {\LARGE Effect of charge on negative--phase--velocity
propagation of electromagnetic waves in the ergosphere of a rotating
black hole
 }}

 \vspace{10mm} \large

Benjamin M. Ross\footnote{Permanent address: Department of
Engineering Science and Mechanics, Pennsylvania State University,
University Park, PA 16802--6812, USA. email: bmr180@psu.edu}, Tom G.
Mackay\footnote{ e--mail: T.Mackay@ed.ac.uk} \\
{\em School of Mathematics,
University of Edinburgh, Edinburgh EH9 3JZ, UK}\\
\bigskip

 Akhlesh  Lakhtakia\footnote{ e--mail: akhlesh@psu.edu}\\
 {\em Department of Engineering Science and
Mechanics\\ Pennsylvania State University, University Park, PA
16802--6812, USA}\\

\end{center}

\vspace{4mm} \normalsize

\begin{abstract}
\noindent The presence of charge promotes the propensity of a rotating
black hole to support the propagation of electromagnetic plane
waves with negative phase velocity (NPV)  in its
ergosphere, whether
the Kerr--Newman or the Kerr--Sen metric
descriptions of spacetime  are
considered.   Striking  differences in the NPV
characteristics for the Kerr--Newman and Kerr--Sen metrics emerge
from numerical studies, particularly close to the outer event horizon when
the magnitude of the charge is large.

\end{abstract}


\section{Introduction}

Electromagnetic plane waves with negative phase velocity (NPV) have
been the centre of much attention lately \c{LMW03}. They give rise
to many exotic phenomenons~---~the most prominent being  negative
refraction \c{Rama}~---~which are not associated with their positive
phase velocity (PPV) counterparts. Much of the interest in NPV
propagation has stemmed from the realization of negatively
refracting metamaterials \c{SSS}. In addition,  the prospects of NPV
propagation in special \c{ML04a} and general \c{MLS05} relativistic
scenarios, and the concomitant implications for observational
astronomy, have  prompted study and speculation within the past few
years \c{Current_Sci}.

It is now well established theoretically that, for certain spacetime metrics, NPV
propagation can be supported in free space \c{LMS05,Schwarzs}. In
particular,   NPV can arise in the
ergosphere of a rotating black hole described by the Kerr metric
\c{Tom}. The NPV--supporting regions are concentrated towards the
equator of the ergosphere, and this concentration becomes
exaggerated as the angular velocity of the black hole is increased. Here
we address the question: how does the presence of charge effect
the propensity of a rotating black hole to support NPV propagation?

The most commonly encountered spacetime description of a charged
rotating black hole is provided by the Kerr--Newman metric
\c{exact}. This solution to the Einstein field equations was derived
in the 1960s \c{Kerr,KerrN}. In 1992, Sen developed an alternative
metric solution describing a charged rotating black hole, based on
the action of a heterotic string \c{KerrS}. Since then, various
aspects of the structure and properties of this Kerr--Sen metric
have been established \c{okai,Blaga,wu}. In the following sections,
we consider both the Kerr--Newman and the Kerr--Sen metrics in our
investigation of the NPV characteristics of a charged rotating black
hole.

\section{Theoretical framework}

\subsection{Preliminaries}

Electromagnetic propagation is considered  in free space. The
spacetime curvature is described by the metric $g_{\alpha
\beta}$\footnote{Greek indexes take the values 0, 1, 2 and 3, Roman
indexes take the values 1, 2 and 3, $x^0 = ct$, where $c$ is the
speed of light in vacuum in the absence of all gravitational fields,
and $x^{1,2,3}$ represent the three spatial coordinates.} with
signature $(+,-,-,-)$. Here we outline the theory of gravitationally
assisted NPV propagation, comprehensive details being available
elsewhere \c{MLS05,Tom}.

 Following the
approach originally put forth by Tamm \c{Skrotskii,Plebanski,LL,SS},
we exploit the fact that electromagnetic propagation in curved
spacetime in free space is equivalent to propagation in a
fictitious, instantaneously responding, bianisotropic medium in flat
spacetime, characterized by the constitutive relations
\begin{equation}
\label{CR2} \left.
\begin{array}{l}
D_\ell = \gamma_{\ell m} E_m + \varepsilon_{\ell mn}\,\Gamma_m\,H_n\\[6pt]
B_\ell =    \gamma_{\ell m} H_m - \varepsilon_{\ell mn}\, \Gamma_m\,
E_n
\end{array}\right\}.
\end{equation}
Herein,
  $E_\ell$, $B_\ell$, $D_\ell$, and $H_\ell$ are the components of the conventional electromagnetic
field vectors; $\varepsilon_{\ell mn}$ is the
three-dimensional Levi-Civita symbol; and the components of the
tensor $\gamma_{\ell m}$ and the vector $\Gamma_m$ are defined as
\begin{equation}
\label{akh1} \left.\begin{array}{l} \gamma_{\ell m}
= \displaystyle{- \le -{g} \ri^{1/2} \, \frac{{g}^{\ell m}}{{g}_{00}}}\\[6pt]
\Gamma_m= \displaystyle{\frac{g_{0m}}{g_{00}}}
\end{array}\ric
\,.
\end{equation}

In order to consider planewave propagation, we rewrite
\r{CR2} in 3$\times$3 dyadic/3 vector form  as
\begin{equation}
\label{eq2}
\left.
\begin{array}{l}
\#D\tr = \epso\,\=\gamma\tr\. \#E\tr - \displaystyle{\frac{1}{c}}\, \#\Gamma\tr\times \#H\tr\,\\
\vspace{-8pt} \\ \#B\tr = \muo\,\=\gamma\tr\. \#H\tr +
\displaystyle{\frac{1}{c}}\,\#\Gamma\tr\times \#E\tr\,
\end{array}
\right\},
\end{equation}
wherein $\=\gamma\tr$ is the dyadic--equivalent of $\gamma_{\ell
m}$,
 $ \#\Gamma\tr$ is the vector--equivalent of $\Gamma_m$,
the scalar constants $\epso$ and $\muo$ denote the permittivity and
permeability of vacuum in the absence of a gravitational field, and
SI units are adopted. In source--free regions, $\#E$, $\#B$, $\#D$,
and $\#H$ are related by the Maxwell curl postulates as
\begin{equation}
\label{eq1} \left.
\begin{array}{l}
\curl \#E\tr + \displaystyle{\frac{\partial}{\partial t}} \#B\tr =
\#0\,\\ \vspace{-8pt} \\ \curl\#H\tr -
\displaystyle{\frac{\partial}{\partial t}} \#D\tr = \#0\,
\end{array}
\right\}.
\end{equation}
Let us  emphasize that, while the spatial ($\#r$) and temporal ($t$)
coordinates have been separated in \r{eq2} and \r{eq1}, the
underlying spacetime remains curved.

Our aim is to explore electromagnetic planewave propagation in the
spacetime of a black hole of geometric mass $m_{bh}$ and charge
$q_{bh}$, rotating with angular momentum $a_{bh}$ about the
Cartesian $z$ axis, in conventional units. Two different
descriptions of this spacetime are considered, as provided by the
Kerr--Newman metric and the Kerr--Sen metric. The line elements for
these are conventionally expressed in terms of Boyer--Lindquist
coordinates. In order to establish the corresponding forms of the
dyadic $\=\gamma$ and vector $\#\Gamma$, the Kerr--Schild Cartesian
coordinate representations for these metrics are presented in the
following two subsections.

\subsection{Kerr--Newman spacetime}
\label{KN}

The Kerr--Newman line element is given in Boyer--Lindquist
coordinates as \c{Bose, Xulu}
\begin{eqnarray}
ds^2 &=& \frac{\Dkn}{R^2 + \arbh^2 \cos^2 \theta} \le dt - \arbh
\sin^2 \theta d \phi \ri^2 - \frac{\sin^2 \theta}{R^2 + \arbh^2
\cos^2
\theta} \les \le R^2 + \arbh^2 \ri d \phi - \arbh dt \ris^2 \nonumber \\
&& - \frac{R^2 + \arbh^2 \cos^2 \theta}{\Dkn} dR^2 - \le R^2 +
\arbh^2 \cos^2 \theta \ri d \theta^2 \l{KN_le},
\end{eqnarray}
where $R$ is given implicitly via
\begin{equation}
R^2 = x^2 + y^2 + z^2 - \arbh^2 \les 1 - \le \frac{z}{R} \ri^2
\ris\, \l{R_def},
\end{equation}
and
\begin{equation}
\Dkn = R^2 - 2 \mrbh R + \arbh^2 + \qrbh^2 \,.
\end{equation}
By applying the coordinate transformation $\le t, \phi, R, \theta
\ri$ $\longrightarrow$ $\le \overline{t}, x, y, z \ri$ specified by
\begin{eqnarray}
x &=& \le R \cos \overline{\phi} + \arbh \sin \overline{\phi} \ri \sin \theta\,, \l{x_coord}\\
y &=& \le R \sin \overline{\phi} - \arbh \cos \overline{\phi} \ri \sin \theta\,,\\
z &=& R \cos \theta\,,\\
d\overline{t}& =& dt + dR - \frac{R^2 + \arbh^2}{\Dkn} d R\,, \l{t_coord}\\
d\overline{\phi}& =& d \phi - \frac{\arbh}{\Dkn} d R\,
\l{phi_coord},
\end{eqnarray}
the line element \r{KN_le} is expressed in Kerr--Schild Cartesian
form as
\begin{eqnarray}
ds^2 &=&
d\overline{t}^2 - dx^2 - dy^2 - dz^2 - \frac{\le 2 \mrbh R - \qrbh^2 \ri R^2}{R^4 + \arbh^2 z^2 } \times \nonumber \\
&& \les d\overline{t} - \frac{z}{R}dz - \frac{R}{R^2 + \arbh^2} \le
x dx + y dy \ri - \frac{\arbh}{R^2 + \arbh^2} \le x dy - y dx \ri
\ris^2 \,. \l{KN_le_KS}
\end{eqnarray}
The corresponding metric components $g_{\alpha \beta}$~---~from
which the associated dyadic $\=\gamma$ and vector $\#\Gamma$ are
immediately delivered by the definitions \r{akh1}~---~are provided
explicitly in Appendix~1.

We focus on the regime $\mrbh^2 > \arbh^2$. Furthermore, our
interest lies in the ergosphere demarcated as $R_+ < R < R_{S+}$,
with $R=R_+$ representing the outer event horizon and $R= R_{S+}$
representing the stationary limit surface \c{Inverno}. The outer
event horizon occurs where $\Dkn = 0$. Thus,
\begin{equation}
R_+ = \mrbh +\sqrt{\mrbh^2-\arbh^2-\qrbh^2}.
\end{equation}
The stationary limit surface $R = R_{S_+}$ occurs where $g_{00} =
0$. Thus,
\begin{equation}
R_{S_+}  = \mrbh +\sqrt{\mrbh^2-\qrbh^2- \le \frac{\arbh z}{R_{S+}}
\ri^2}.
\end{equation}

\subsection{Kerr--Sen spacetime} \l{KS}

We now turn to the Kerr--Sen  line element which is given in
Boyer--Lindquist coordinates as \c{okai,wu}
\begin{eqnarray}
ds^2 &=& \frac{\Dks}{\tilde{S}^2 + \arbh^2 \cos^2 \theta} \le dt -
\arbh \sin^2 \theta d \phi \ri^2 - \frac{\sin^2 \theta}{\tilde{S}^2
+ \arbh^2 \cos^2
\theta} \les \le \tilde{S}^2 + \arbh^2 \ri d \phi - \arbh dt \ris^2 \nonumber \\
&& - \frac{\tilde{S}^2 + \arbh^2 \cos^2 \theta}{\Dks} dS^2 - \le
\tilde{S}^2 + \arbh^2 \cos^2 \theta \ri d \theta^2, \l{KS_le}
\end{eqnarray}
where
\begin{eqnarray}
\Dks &=& S^2  - 2 \le \mrbh - \frac{\beta}{2} \ri  S
+  \arbh^2\,,\\
\tilde{S}^2 &=& S^2 + \beta S\,, \l{S_tilde}\,\\
\beta &=& \frac{\qrbh^2}{\mrbh}\,,
\end{eqnarray}
and $\tilde{S}$ satisfies
\begin{equation}
\tilde{S}^2 = x^2 + y^2 + z^2 - \arbh^2 \les 1 - \le
\frac{z}{\tilde{S}} \ri^2 \ris\, \l{ST_def}.
\end{equation}
Under the  coordinate transformation $\le t, \phi, S, \theta \ri$
$\longrightarrow$ $\le \overline{t}, x, y, z \ri$ specified by
\begin{eqnarray}
x &=& \le \tilde{S} \cos \overline{\phi} + \arbh \sin \overline{\phi} \ri \sin \theta\,, \l{xks_coord}\\
y &=& \le \tilde{S} \sin \overline{\phi} - \arbh \cos \overline{\phi} \ri \sin \theta\,,\\
z &=& \tilde{S} \cos \theta\,,\\
d\overline{t}& =& dt + d\tilde{S} - \frac{\tilde{S}^2 + \arbh^2}{\Dks} d \tilde{S}\,, \l{tks_coord}\\
d\overline{\phi}& =& d \phi - \frac{\arbh}{\Dks} d \tilde{S}\,
\l{phiks_coord},
\end{eqnarray}
the Kerr--Schild Cartesian form of the line element \r{KS_le}
emerges as
\begin{eqnarray}
ds^2 &=& d\overline{t}^2 - dx^2 - dy^2 - dz^2 + \frac{\beta^2 \le
\tilde{S}^2 + \arbh^2 \cos^2 \theta \ri}{\le \beta^2 + 4 \tilde{S}^2
\ri \Dks} d\tilde{S}^2
- \frac{ 2 \mrbh \tilde{S}^2  S}{\tilde{S}^4 + \arbh^2 z^2 } \times \nonumber \\
&& \les d\overline{t} - \frac{z}{\tilde{S}}dz -
\frac{\tilde{S}}{\tilde{S}^2 + \arbh^2} \le x dx + y dy \ri -
\frac{\arbh}{\tilde{S}^2 + \arbh^2} \le x dy - y dx \ri \ris^2 \,.
\l{KS_le_KS}
\end{eqnarray}
In Appendix~2, explicit expressions for the corresponding metric
components $g_{\alpha \beta}$ are presented. From these, the
associated dyadic $\=\gamma$ and vector $\#\Gamma$ immediately
delivered by the definitions \r{akh1}.

 In analogous manner to the case for the Kerr--Newman
metric, the outer event horizon  arises at $S = S_+$ with
\begin{equation}
 S_+  = \mrbh - \frac{\beta}{2} + \sqrt{ \le \mrbh -
\frac{\beta}{2} \ri^2 - \arbh^2 } \,,
\end{equation}
whereas the stationary limit surface  arises at $S =S_{S_+}$ with
\begin{equation}
S_{S_+} =  \mrbh - \frac{\beta}{2} + \sqrt{ \le \mrbh -
\frac{\beta}{2}\ri^2 - \le \frac{ \arbh z}{ \tilde{S}_{S_+} }\ri^2 }
\,.
\end{equation}
As in \S\ref{KN},  the ergosphere is the region  bounded by the
outer event horizon and the stationary limit surface; i.e., $ S_+ <
S < S_{S_+}$.

\subsection{Planewave propagation}

Now we turn to the propagation of plane waves in the medium
characterized by \r{eq2}. By choosing a sufficiently small
neighbourhood $\cal R$ specified by the Cartesian coordinates $\le
\tilde{x}, \tilde{y}, \tilde{z} \ri$, we approximate the non-uniform
metric $g_{\alpha \beta}$ with the uniform metric ${\tilde
g}_{\alpha \beta}$ throughout  $\cal R$ \c{LMS05}. Within the
neighbourhood $\cal R$, we utilize the uniform 3$\times$3 dyadic $
\={\tilde\gamma} \equiv \left. \=\gamma \,\right|_{\cal R}$ and the
uniform 3 vector $ \#{\tilde\Gamma} \equiv \left. \#\Gamma \,
\right|_{\cal R} $, where $ \={\tilde\gamma}$ and $
\#{\tilde\Gamma}$ are calculated from \r{akh1}. Due to the  uniform
approximation, attention is restricted  to electromagnetic
wavelengths that are small compared to the linear dimensions of
$\cal R$.

The three--dimensional Fourier transforms
\begin{equation}
\left.
\begin{array}{l}
\#E (ct, \#r) = \displaystyle{ \frac{1}{c} \int_{-\infty}^\infty
\int_{-\infty}^\infty \int_{-\infty}^\infty
 \#{\sf E}(\omega/c, \#k) \exp\les i(\#k \. \#r - \omega t)\ris \, d\omega \, dk_1  \, dk_2\,} \\ \vspace{-2mm} \\
\#H (ct, \#r)  = \displaystyle{ \frac{1}{c} \int_{-\infty}^\infty
\int_{-\infty}^\infty  \int_{-\infty}^\infty
 \#{\sf H} (\omega/c, \#k) \exp\les i(\#k \. \#r - \omega t )\ris \, d \omega \, dk_1 \, dk_2\, }
\end{array}
\right\}, \label{der}
  \end{equation}
 decompose the electromagnetic fields within $\cal R$, Here,  the
wavevector $\#k$ is the Fourier variable corresponding to $\#r$,
$i=\sqrt{-1}$ and $\omega$ is the usual temporal frequency. By
combining  \r{eq1} and \r{der}, and then  solving the resulting 4
$\times$ 4 eigenvalue problem, the wavevector component $k_3$ is
determined \c{Lopt92}. Only propagating (non--evanescent) plane
waves are considered; i.e.,  $k_3 \in \mathbb{R}$.

NPV propagation occurs when the phase velocity vector casts a negative
projection on to the time-averaged Poynting vector~---~i.e.,
\begin{equation}
\#k \. \langle \, \#{\sf P} (\omega/c, \#k) \, \rangle_t < 0\,,
\label{cdef}
\end{equation}
where the time-averaged Poynting vector $\langle \, \#{\sf P} (\omega/c, \#k) \, \rangle_t $ is calculated from the
complex--valued phasors of the electric and magnetic fields,
 $\#{\sf E} (\omega/c, \#k)$ and  $\#{\sf H} (\omega/c, \#k)$.
 The  left side of \r{cdef}  is given by \c{Tom}
\begin{eqnarray}
\#k \. \langle \#{\sf P} (\omega/c, \#k) \rangle_t &=& \frac{1}{2
\omega \muo \vert \={\tilde\gamma} \vert} \le \vert A_a \vert^2
\#{\sf e}_a \. \={\tilde\gamma} \. \#{\sf e}_a +  \vert A_b \vert^2
\#{\sf e}_b \. \={\tilde\gamma} \. \#{\sf e}_b \ri \, \#k \.
\={\tilde\gamma} \. \#p \,. \label{ppp1}
\end{eqnarray}
The complex--valued amplitudes $
A_{a,b} \ok$ are fixed by initial and boundary conditions, whereas
the unit vectors
\begin{equation}
\left.
\begin{array}{l}
 \#{\sf e}_a
= \displaystyle{\frac{\={\tilde\gamma}^{-1} \. \#w}
{\vert\={\tilde\gamma}^{-1} \. \#w\vert}} \\ \\\  \#{\sf e}_b =
\displaystyle{\frac{\={\tilde\gamma}^{-1}\. \le \#p \times \#{\sf
e}_a \ri}{\vert\={\tilde\gamma}^{-1}\. \le \#p \times \#{\sf e}_a
\ri\vert}}
\end{array}
\right\}
 \,. \l{e_12}
\end{equation}
 For the numerical results presented in the
following section, we have chosen
\begin{equation}
\#w = \left\{ \begin{array}{lcr} \displaystyle{\frac{1}{\sqrt{2
p^2_z + \le p_x + p_y \ri^2} } }\le p_z, p_z, -p_x-p_y \ri &
\mbox{for} & p_z \neq 0
\\ & \\
\displaystyle{ \frac{1}{\sqrt{ p_y^2 + p^2_x}}} \le p_y, -p_x, 0 \ri
& \mbox{for} & p_z = 0
\end{array}
\right. ,
\end{equation}
with $\#p\equiv \le p_x, p_y, p_z \ri$ in the Cartesian coordinate
system.

For both the Kerr--Newman and Kerr--Sen metrics  $\vert \,
\={\tilde\gamma}\, \vert
> 0$. Hence, the sufficient condition for NPV propagation
\begin{equation} \l{NPV_cond}
\left.
\begin{array}{l}
P_a < 0 \\
P_b < 0
\end{array}
\right\}
\end{equation}
arises, wherein
\begin{equation}
\left. \l{PaPb}
\begin{array}{l}
P_a = \le \, \#{\sf e}_a \. \={\tilde\gamma} \. \#{\sf e}_a \, \ri
\le \, \#k \. \={\tilde\gamma} \. \#p \, \ri \\ \vspace{-6pt}
\\ P_b = \le \, \#{\sf e}_b \. \={\tilde\gamma} \. \#{\sf e}_b ,\
\ri \le \, \#k \. \={\tilde\gamma} \. \#p \, \ri
\end{array}
\right\}.
\end{equation}
The two wavenumbers $k = k^\pm$ given by
\begin{equation} \l{wave_nos}
k^\pm = \frac{\omega}{c} \le \frac{\hat{\#k} \. \={\tilde\gamma} \.
\#{\tilde \Gamma} \pm \sqrt{ \le \, \hat{\#k} \. \={\tilde\gamma} \.
\#{\tilde \Gamma} \, \ri^2 - \hat{\#k} \. \={\tilde\gamma} \.
\hat{\#k} \le \#{\tilde\Gamma} \. \={\tilde\gamma} \.
\#{\tilde\Gamma} - \vert \,\={\tilde\gamma} \, \vert\,
\ri}}{\hat{\#k} \. \={\tilde\gamma} \. \hat{\#k}}\ri
\end{equation}
develop for the arbitrarily oriented wavevector $\#k = k \hat{\#k}$
with $\hat{\#k} = \le \sin \theta \cos \phi, \sin \theta \sin \phi,
\cos \theta \ri$.

\section{Numerical results}

The NPV characteristics of an uncharged rotating black hole, as
described by the Kerr metric, have been established previously
\c{Tom}. In this section, we report on  the effect that charge has
on the propensity for a rotating black hole to support NPV
propagation in its ergosphere. This study was carried out by
numerically evaluating \r{ppp1}, for both the Kerr--Newman and
Kerr--Sen descriptions of the blackhole spacetime.

The following strategy was implemented: for fixed values of
$\mrbh$, $\arbh$ and $\qrbh$, we considered surfaces of constant $R$
for the Kerr--Newman metric, and surfaces of constant $S$ for the
Kerr--Sen metric, with all surfaces being contained within the
respective blackhole ergospheres. The prospects for NPV propagation for all
propagation directions $\hat{\#k} = \le \sin \theta \cos \phi, \sin
\theta \sin \phi, \cos \theta \ri $ were calculated over a grid of
  points on  surfaces of constant  $R$ and $S$.
 At each coordinate point $\le \tilde{x},
\tilde{y}, \tilde{z} \ri$ on the grid, the quantities $P_a$ and
$P_b$ were computed at $1^\circ$ increments for $ \theta \in
[0^\circ ,180^\circ)$ and $ \phi \in [0^\circ ,360^\circ)$, for both
$k=k^+$ and $k=k^-$. For each value of $( \theta , \phi )$ at which
NPV was found to be  supported (i.e., $P_a < 0$ and $P_b < 0$), a
score of one point was assigned to the $\le \tilde{x}, \tilde{y},
\tilde{z} \ri$ grid point. The total score at each grid point was
translated to a colour for the convenient presentation of results.

\begin{figure}[!ht]
\centering \psfull \epsfig{file=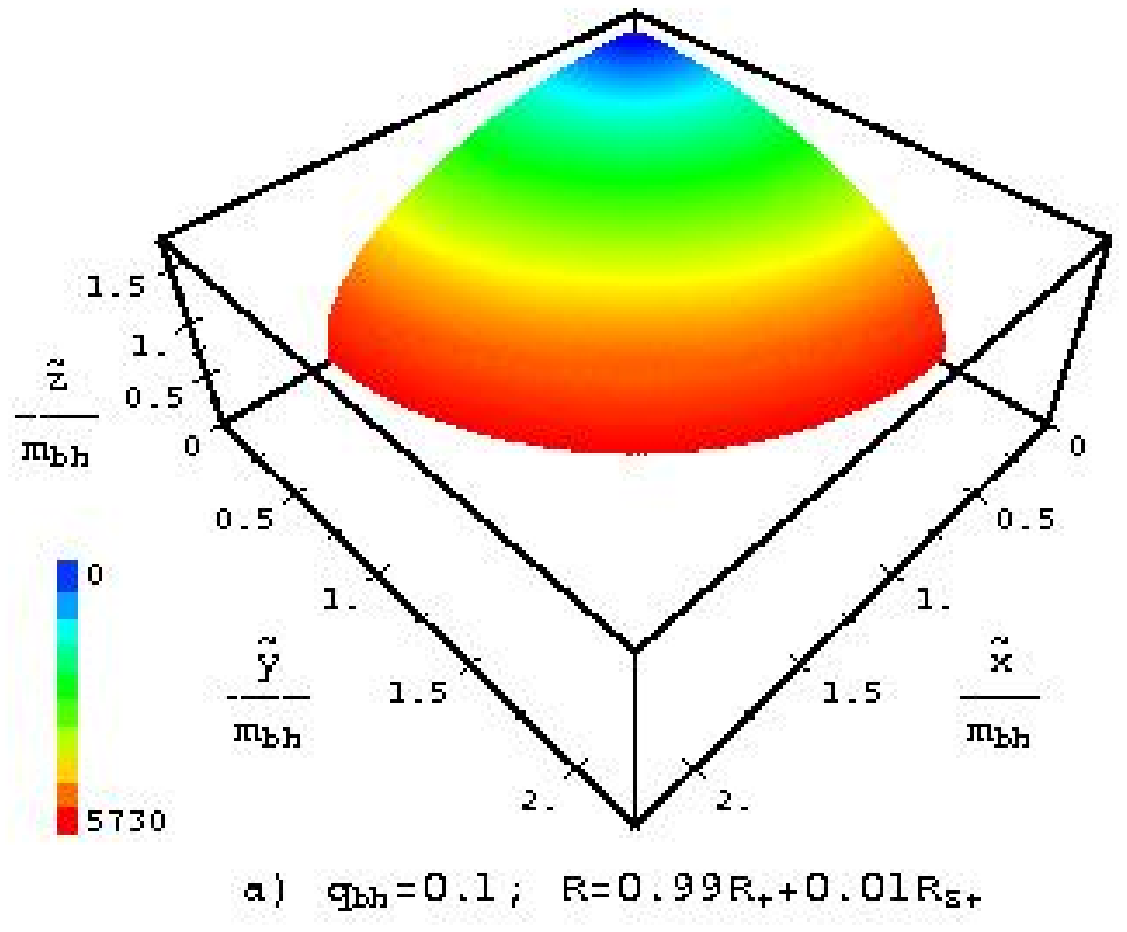,width=3.0in} \hfill
\epsfig{file=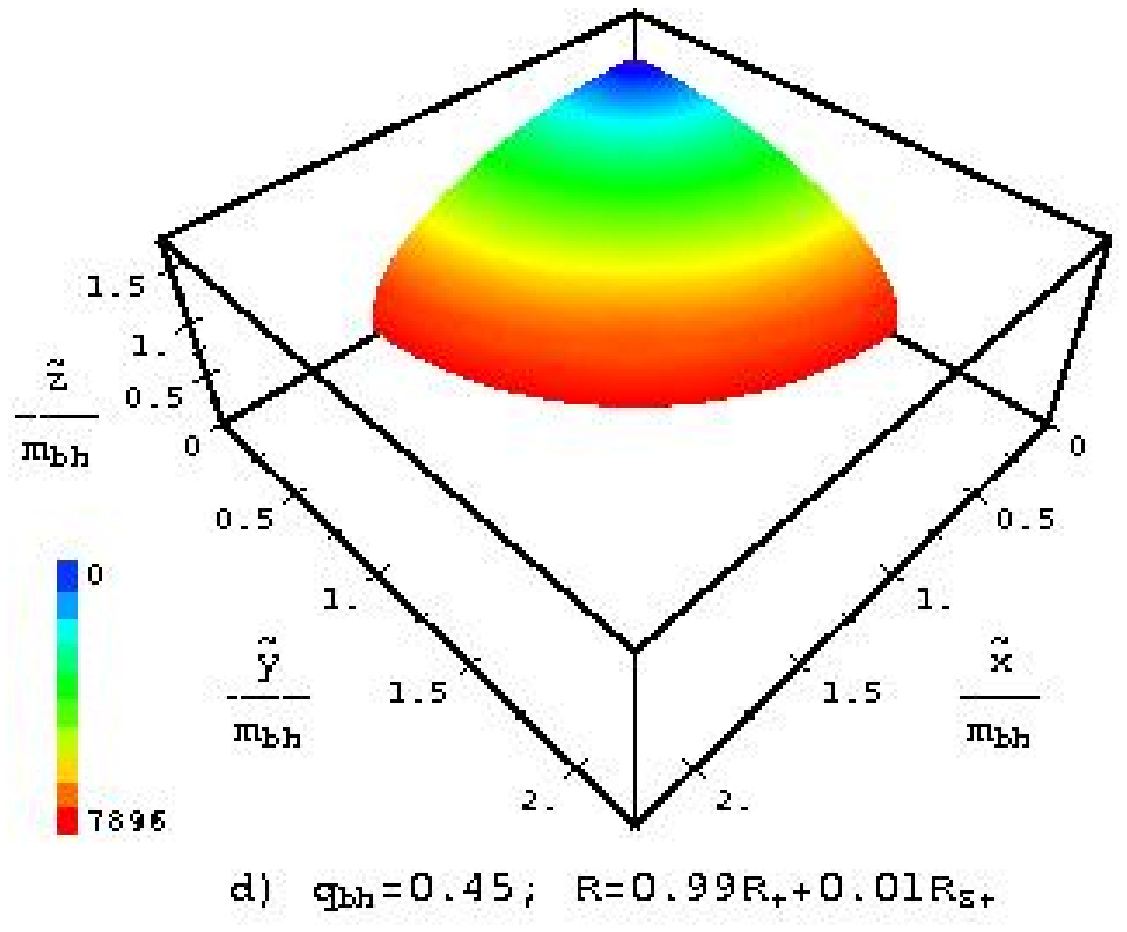,width=3.0in}\\
\epsfig{file=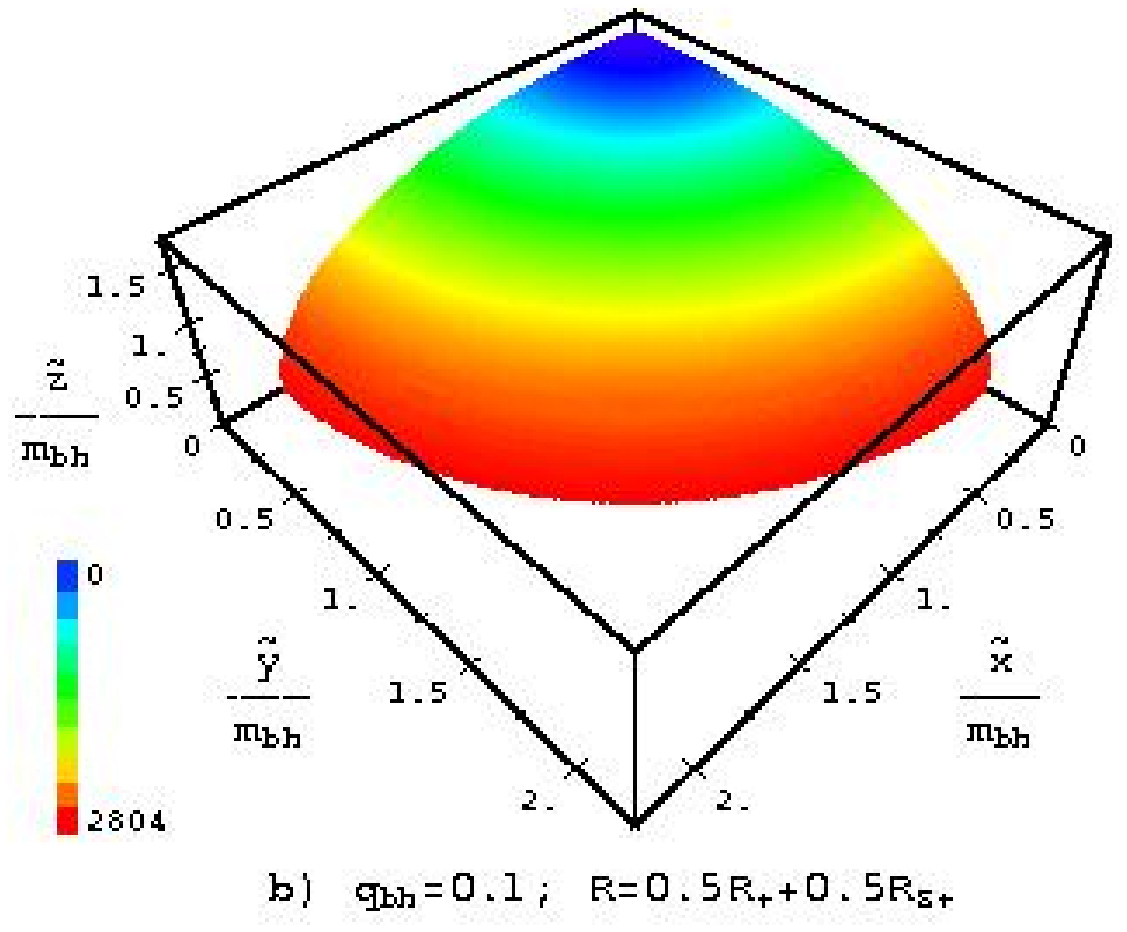,width=3.0in} \hfill
\epsfig{file=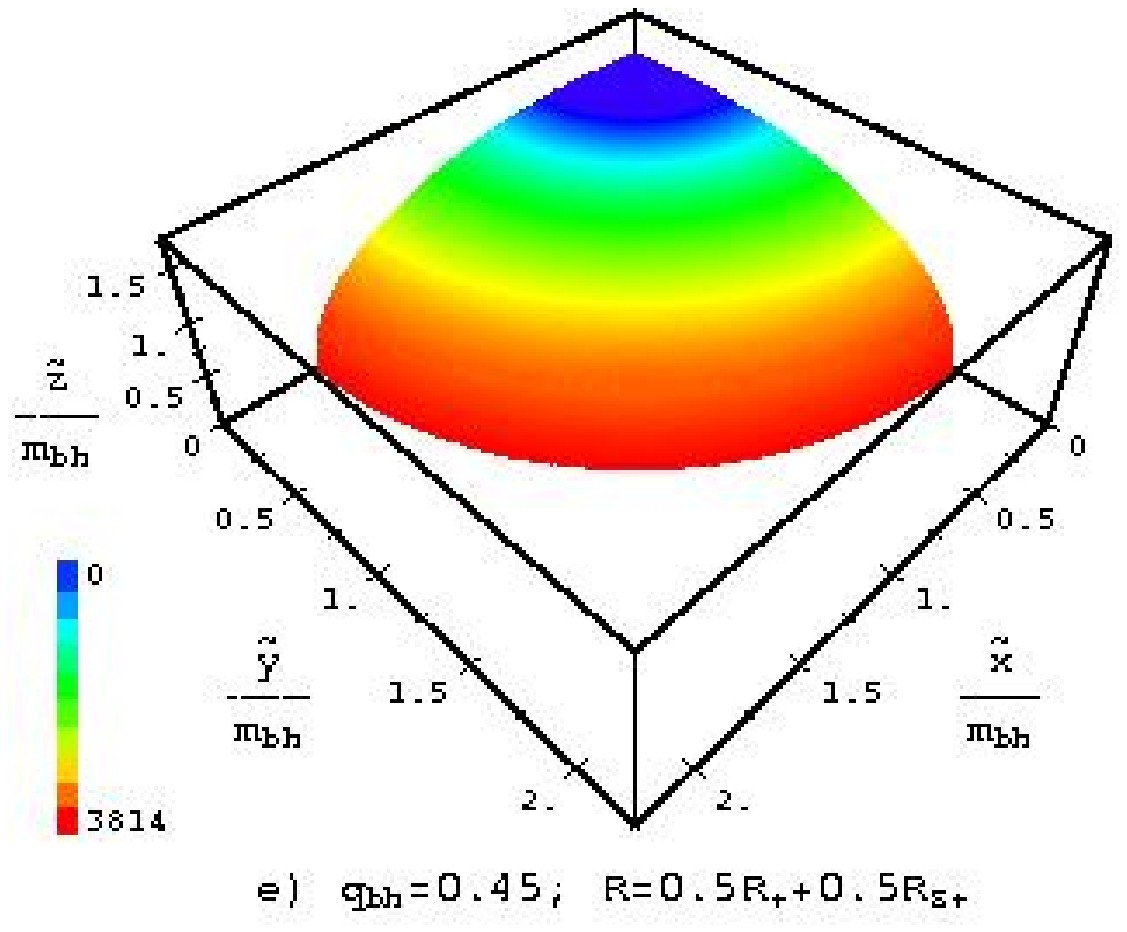,width=3.0in}\\
\epsfig{file=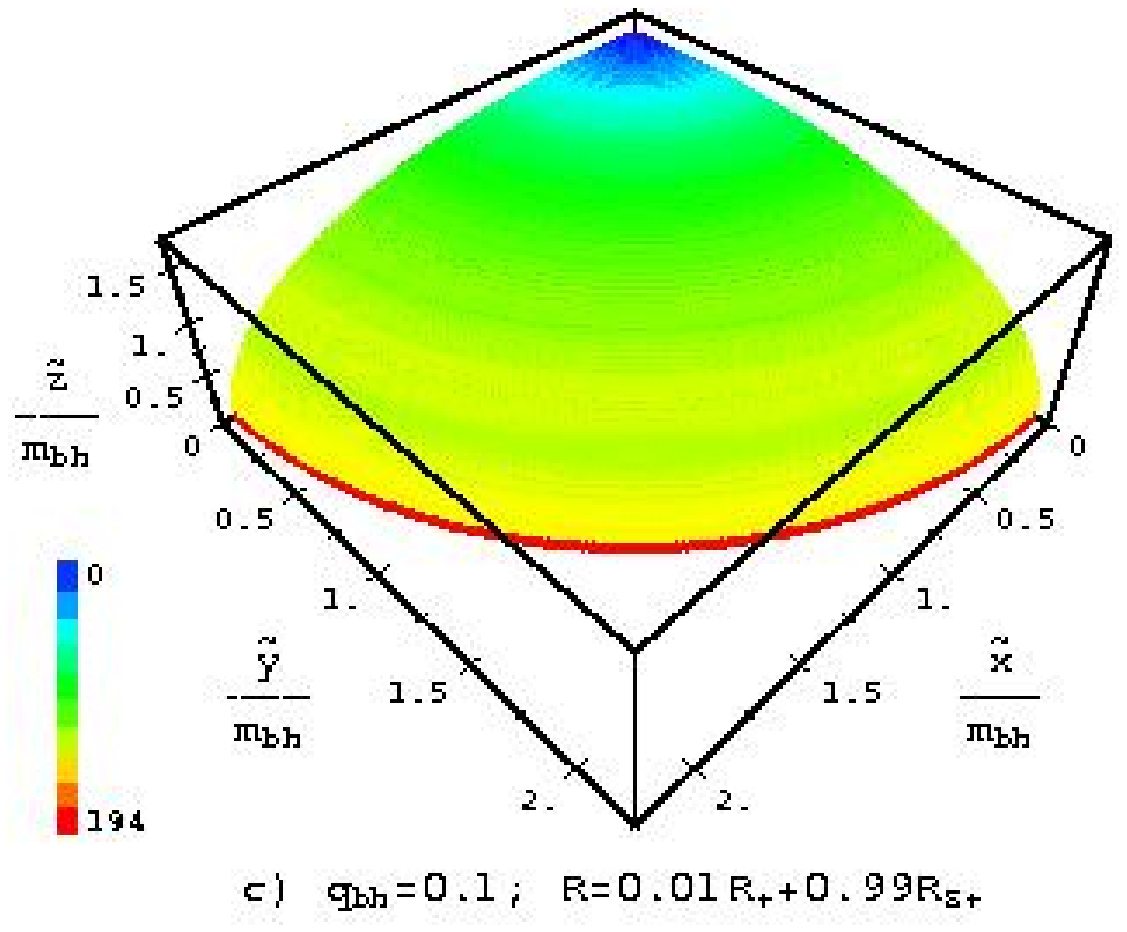,width=3.0in} \hfill
\epsfig{file=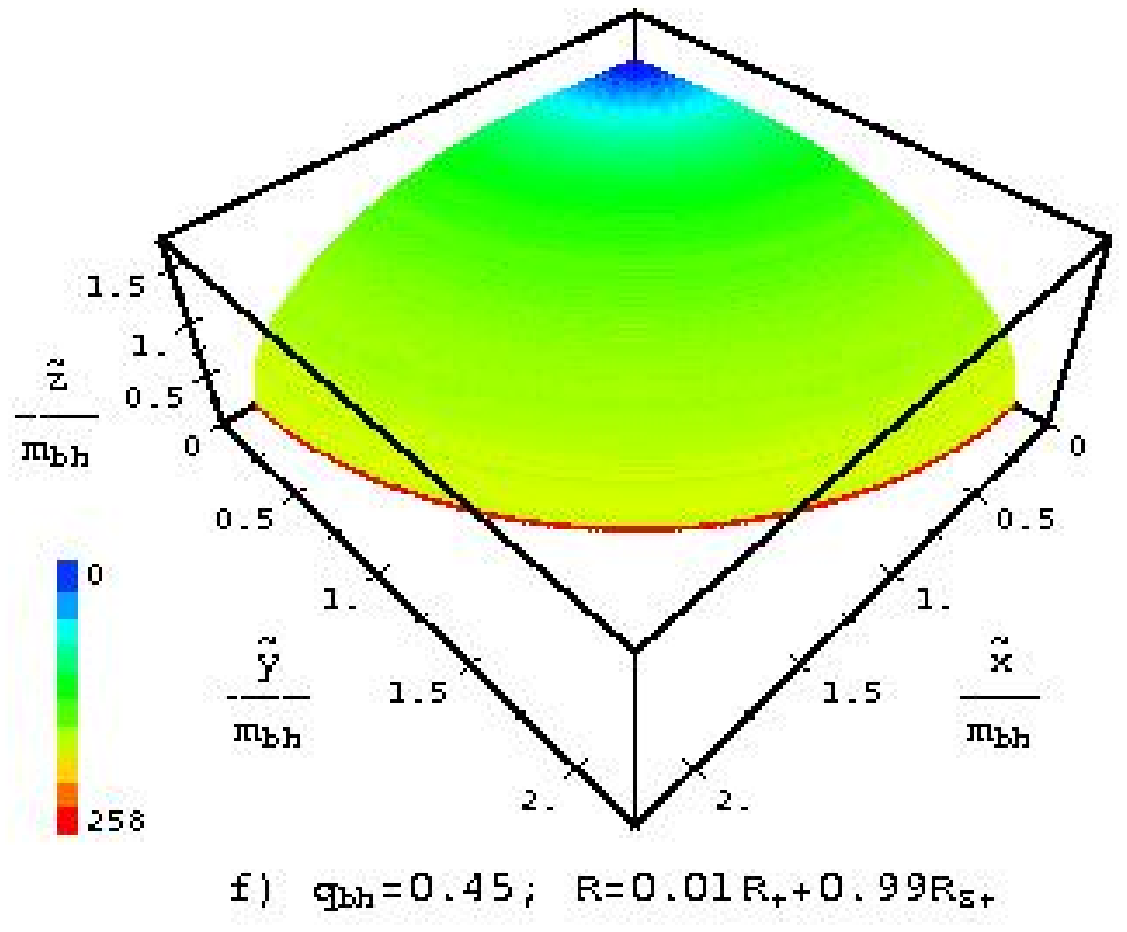,width=3.0in}
 \caption{\label{fig1} NPV maps for $\arbh = \mrbh  \sqrt{3/4}$ for the Kerr--Newman metric.
See the text for an explanation of the colour mapping.
 (a) At the surface $R = 0.99 R_{+} + 0.01
 R_{S_+}$ with $\qrbh=0.1$; (b) at the surface $R = 0.5 R_{+} + 0.5
 R_{S_+}$ with $\qrbh=0.1$; (c) at the surface $R = 0.01 R_{+} + 0.99
 R_{S_+}$ with $\qrbh=0.1$; (d) at the surface $R = 0.99 R_{+} + 0.01
 R_{S_+}$ with $\qrbh=0.45$; (e) at the surface $R = 0.5 R_{+} + 0.5
 R_{S_+}$ with $\qrbh=0.45$; and (f) at the surface $R = 0.01 R_{+} + 0.99
 R_{S_+}$ with $\qrbh=0.45$.
  }
\end{figure}

Figure~\ref{fig1}  shows the NPV propagation maps for Kerr--Newman
spacetime for a charged rotating black hole with $\arbh = \mrbh
\sqrt{3/4}$.
 Three constant $R$ surfaces within
the ergosphere are considered, namely $R= 0.99 R_+ + 0.01 R_{S_+}$
(in Figures~\ref{fig1}(a) and (d)); $R= 0.5 R_+ + 0.5 R_{S_+}$ (in
Figures~\ref{fig1}(b) and (e)); and  $R= 0.01 R_+ + 0.99 R_{S_+}$
(in Figures~\ref{fig1}(c) and (f)). For each surface, a  small value
of $\qrbh$ ($\qrbh = 0.1$ in Figure~\ref{fig1}(a), (b), and (c)) and
a large value of $\qrbh$ ($\qrbh = 0.45$ in Figure~\ref{fig1}(d),
(e), and (f)) are chosen. We note that  $\qrbh \leq 0.5$
in order for the ergosphere boundaries to remain real--valued.

In Figure~\ref{fig1}(a)--(c) that, for $\qrbh = 0.1$, the
regions of NPV propagation are most heavily concentrated near the
equator of the ergosphere and NPV becomes less likely towards the
poles. Also, NPV propagation is more likely near the outer event
horizon and less likely near the stationary limit surface.
Increasing the value of $\qrbh$ does not markedly change the
qualitative nature of the NPV characteristics, as illustrated by
Figure~\ref{fig1}(d)--(f). However, the incidence of NPV propagation
is considerably greater at  $\qrbh = 0.45$ as compared with $\qrbh =
0.1$, at all three constant $R$ surfaces.

\begin{figure}[!ht]
\centering \psfull \epsfig{file=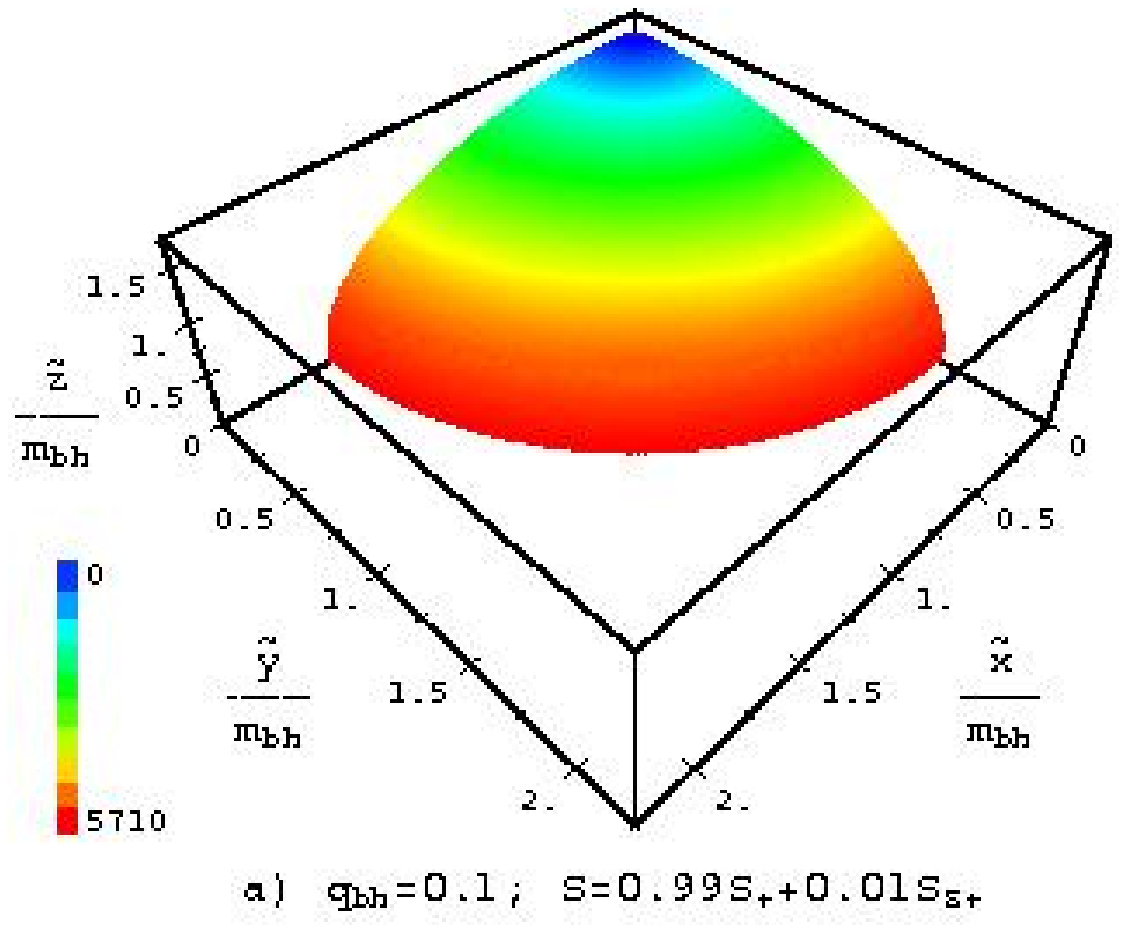,width=3.0in} \hfill
\epsfig{file=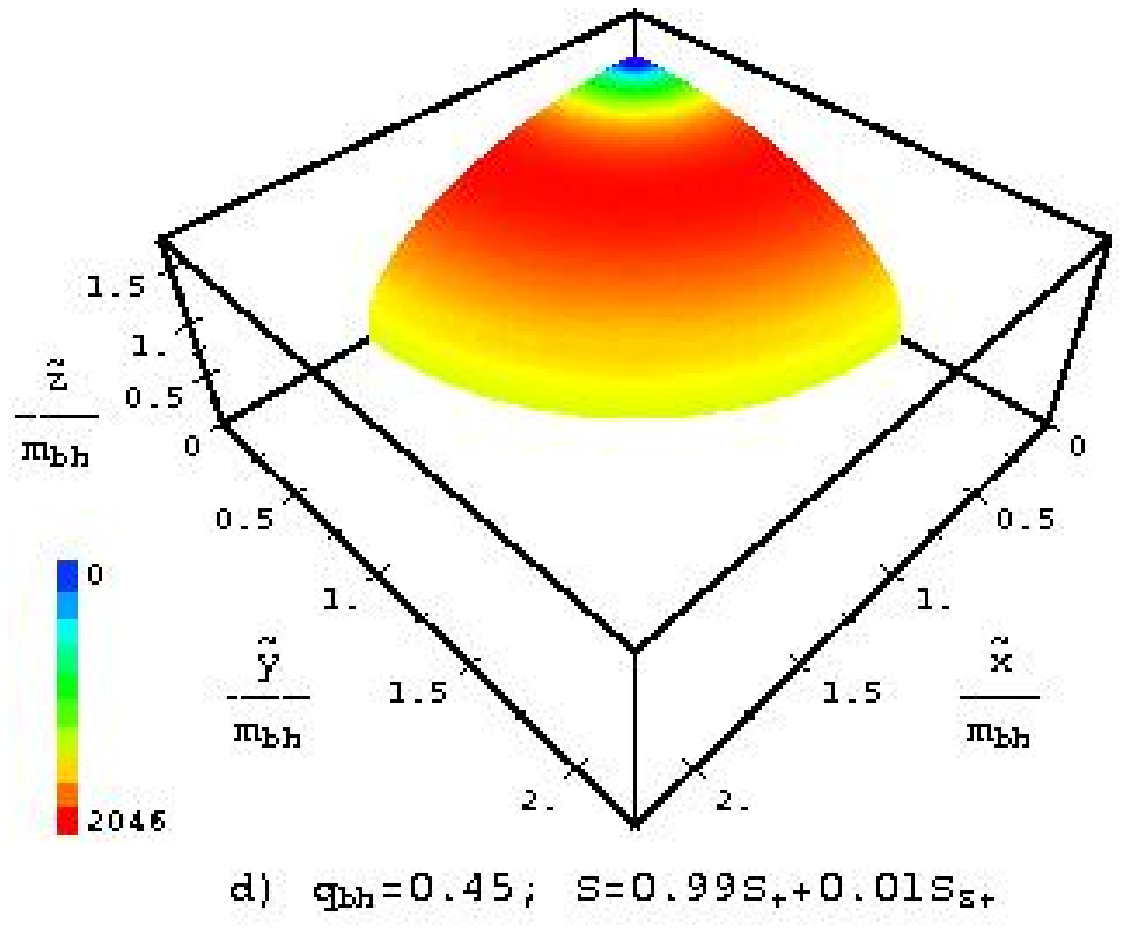,width=3.0in}\\
\epsfig{file=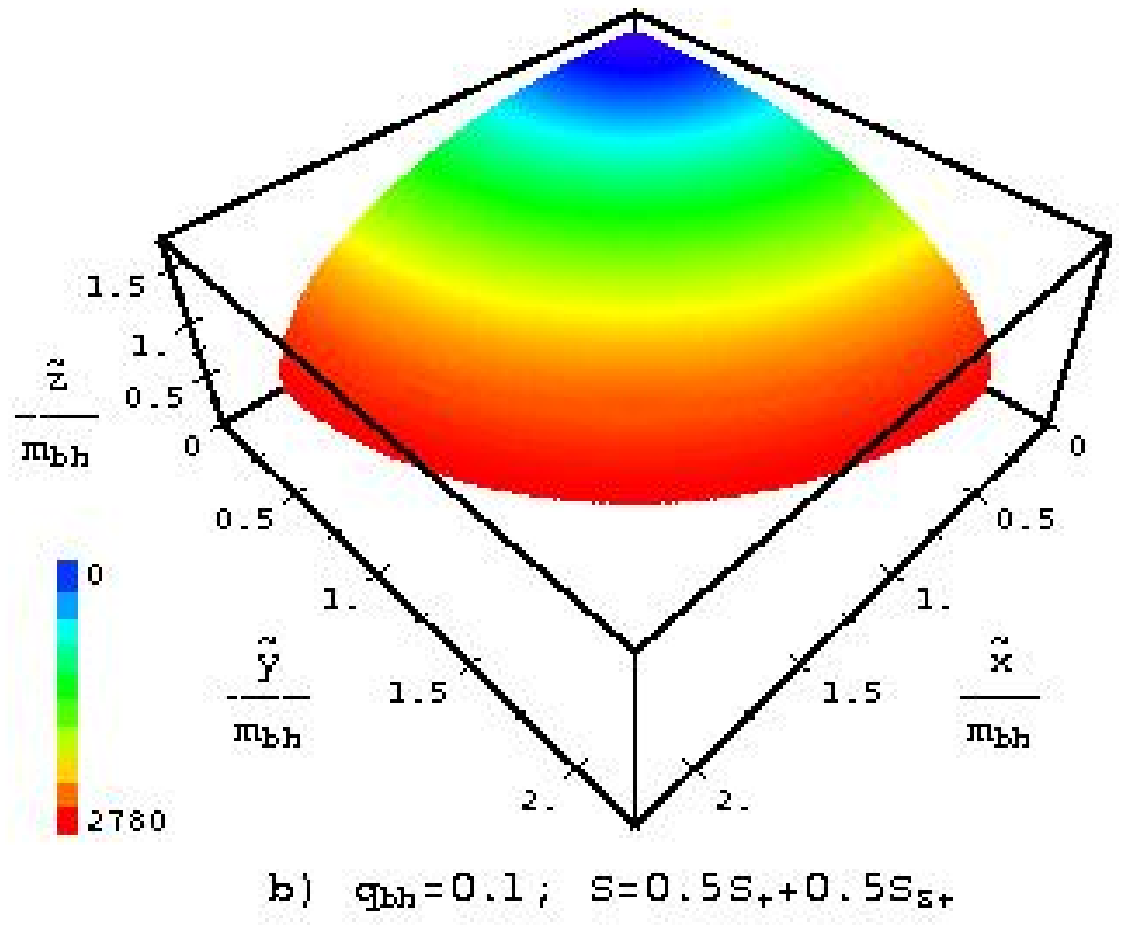,width=3.0in} \hfill
\epsfig{file=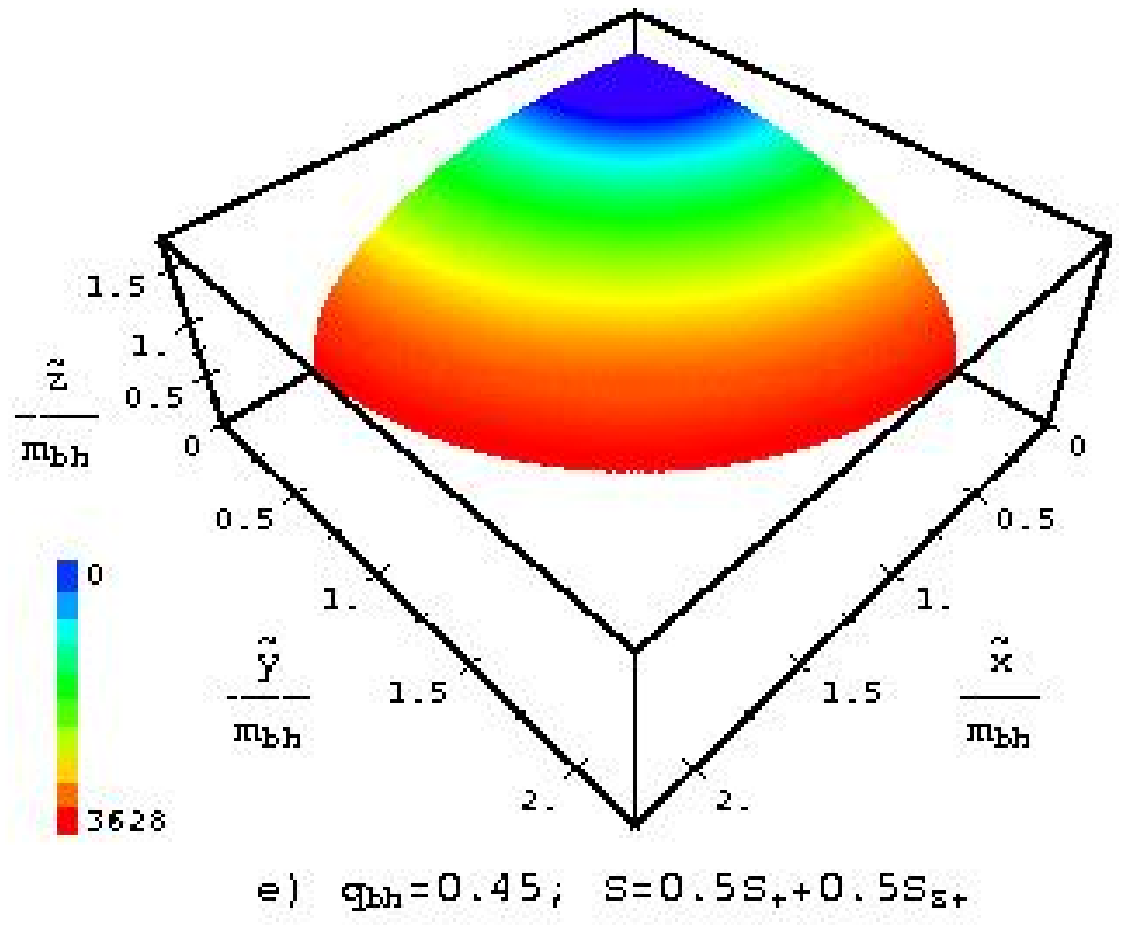,width=3.0in}\\
\epsfig{file=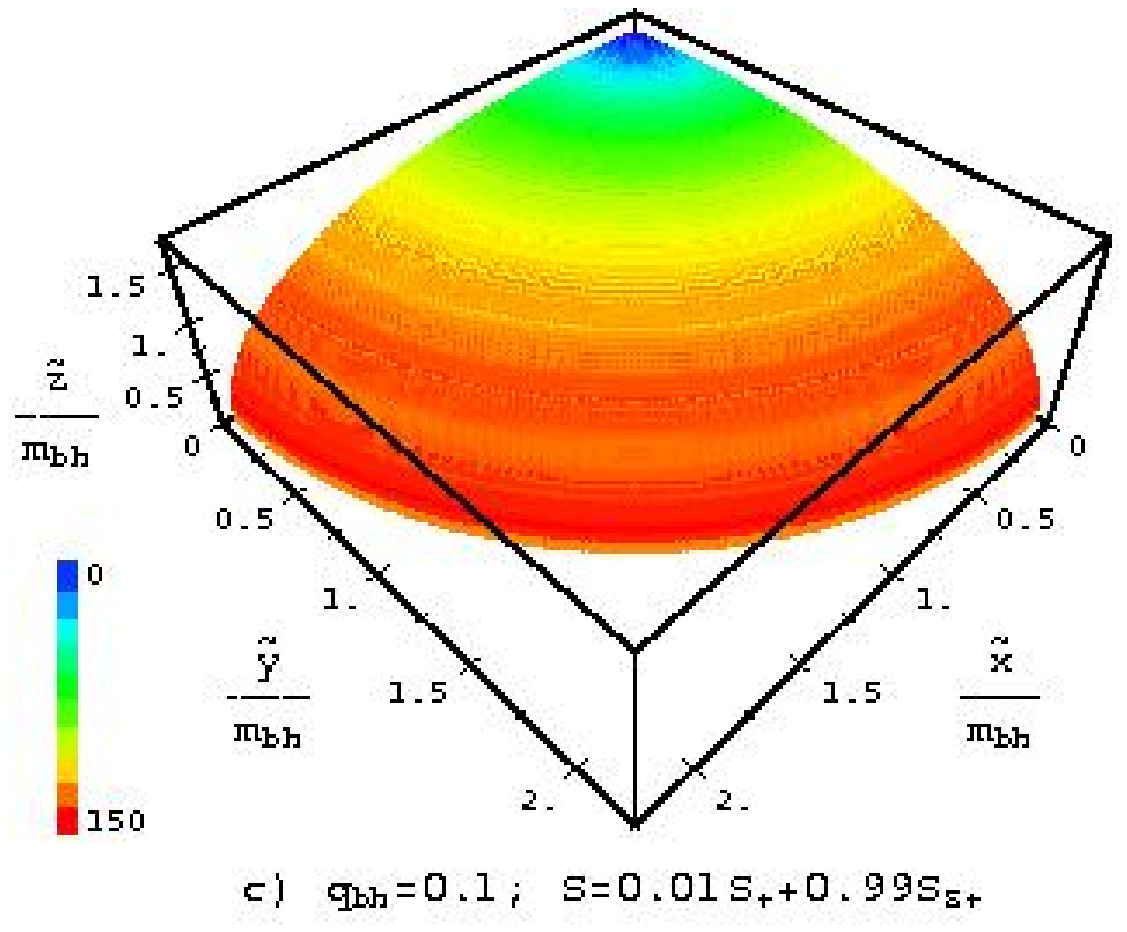,width=3.0in} \hfill
\epsfig{file=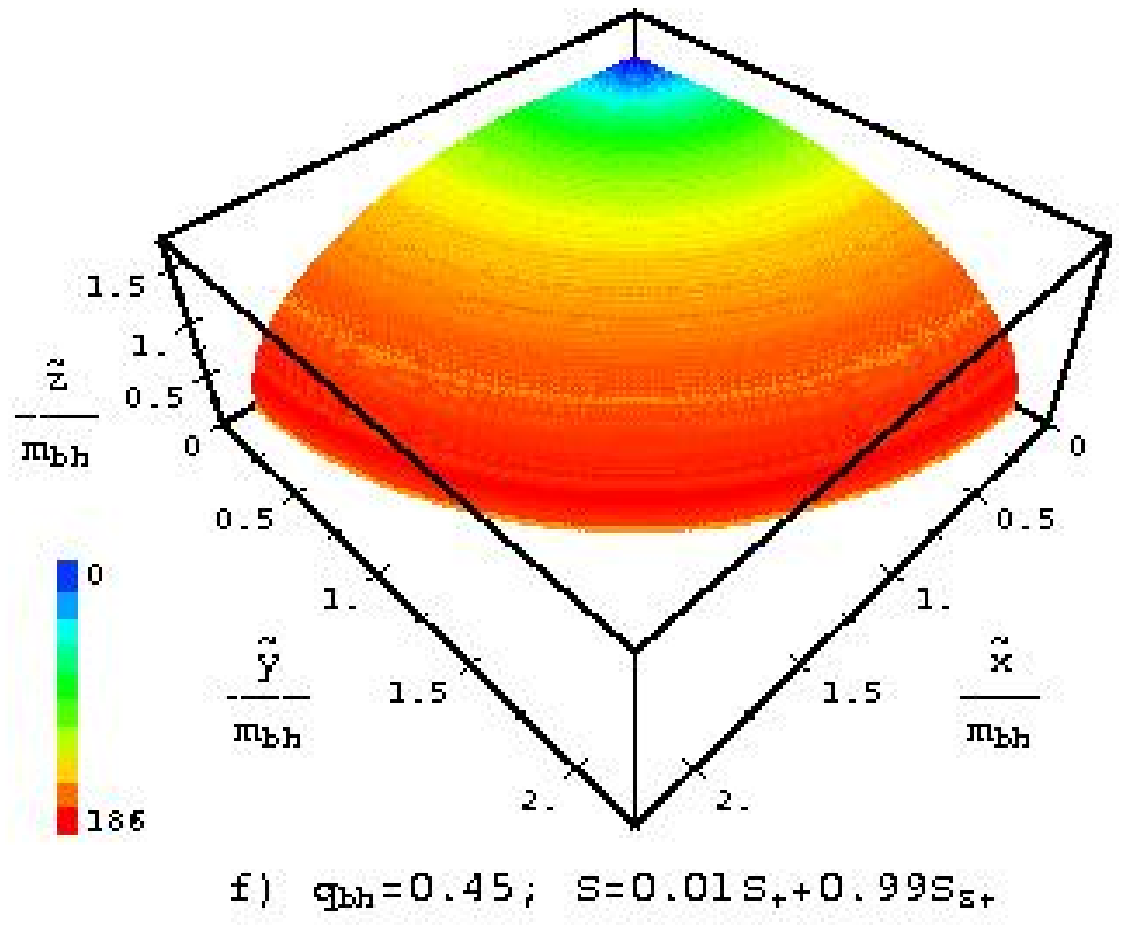,width=3.0in}
 \caption{\label{fig2}
 NPV maps for $\arbh = \mrbh  \sqrt{3/4}$ for the Kerr--Sen metric.
See the text for an explanation of the colour mapping.
 (a) At the surface $S = 0.99 S_{+} + 0.01
 S_{S_+}$ with $\qrbh=0.1$; (b) at the surface $S = 0.5 S_{+} + 0.5
 S_{S_+}$ with $\qrbh=0.1$; (c) at the surface $S = 0.01 S_{+} + 0.99
 S_{S_+}$ with $\qrbh=0.1$; (d) at the surface $S = 0.99 S_{+} + 0.01
 S_{S_+}$ with $\qrbh=0.45$; (e) at the surface $S = 0.5 S_{+} + 0.5
 S_{S_+}$ with $\qrbh=0.45$; and (f) at the surface $S = 0.01 S_{+} + 0.99
 S_{S_+}$ with $\qrbh=0.45$.
  }
\end{figure}

The corresponding NPV propagation maps calculated using the
Kerr--Sen metric are presented in Figure~\ref{fig2}. A comparison of
Figure~\ref{fig2}(a)--(c) with Figure~\ref{fig1}(a)--(c) reveals
that the NPV characteristics of the Kerr--Sen and Kerr--Newman
metrics are quite similar at $\qrbh = 0.1$. However, stark
differences between the Kerr--Sen and Kerr--Newman NPV
characteristics emerge at $\qrbh = 0.45$, particularly near to the
outer event horizon. For the Kerr--Sen metric, we see in
Figure~\ref{fig2}(d) that the distribution of NPV propagation is
concentrated in a region between the ergosphere pole and equator,
while a relatively low incidence occurs at the equator. This is
quite different from the situation for the Kerr--Newman metric that
is displayed in Figure~\ref{fig1}(d).

We note that the NPV propagation maps at $\qrbh = 0.1$, for both the
Kerr-Newman metric and Kerr--Sen metric, are quite similar to those
reported for the uncharged rotating black hole \c{Tom}. This
reflects the fact that both the Kerr--Newman
 metric and the Kerr--Sen metric become identical to
  the Kerr metric
in the limit $\qrbh \rightarrow 0$. Furthermore, regardless of
charge,  NPV is not supported at the ergosphere pole.

\section{Concluding remarks}

Our investigation has revealed that the propensity for a rotating
black hole to support NPV propagation is highly sensitive to the
presence of charge in the black hole. Specifically,
\begin{itemize}
\item the  overall incidence of NPV  increases as  $\qrbh$ increases, for both
the Kerr--Newman and Kerr--Sen metrics;

\item as $\qrbh$ increases, differences in the NPV characteristics
associated with the  Kerr--Newman   and  Kerr--Sen metrics emerge;
and

\item the greatest differences between the NPV characteristics of the
Kerr--Newman and the Kerr--Sen metrics arise close to the  outer
event horizons of their respective ergospheres.
\end{itemize}
These findings suggest the possibility of
 a NPV--based experiment to
distinguish between the Kerr--Newman and Kerr--Sen metric
descriptions of a charged rotating black hole.

\vspace{20mm}

\noindent{\bf Acknowledgements:}
BMR is supported by a UK EPSRC (grant GR/S60631/01) and a US NSF Graduate Research Fellowship. TGM and AL thank Dr.
Sandi Setiawan (University of Edinburgh) for suggesting this study
and Prof. Malcolm MacCallum (Queen Mary, University of London) for
his assistance with  \S\ref{KS}.

\vspace{5mm} \noindent \Large{{\bf Appendix 1}} \normalsize
\vspace{5mm}

The components of the Kerr--Newman metric obtained from the line
element \r{KN_le_KS} expressed in Kerr--Schild Cartesian form are
\begin{equation}
g_{00}=1-\dkn,
\end{equation}
\begin{eqnarray}
g_{01} = \sigma((-1+\dkn)R^3 x \tau^2+\Dkn\{-\arbh \dkn[R^3(\arbh
x+Ry)+\arbh (-Rx+\arbh y)z^2]+R^3x\tau\}),
\end{eqnarray}
\begin{eqnarray}
g_{02}=\sigma((-1+\dkn)R^3 y \tau^2+\Dkn\{ \arbh \dkn[R^3(Rx-\arbh
y)+ \arbh z^2(\arbh x+Ry)]+R^3y\tau\}),
\end{eqnarray}
\begin{eqnarray}
g_{03}=\frac{z \sigma \tau}{R} \{\Dkn R^4+\arbh^2
\Dkn[-(-1+\dkn)R^2+\dkn z^2]+(-1+\dkn)R^2 \tau^2\},
\end{eqnarray}
\begin{eqnarray}
g_{11}&=&-\sigma^2((-1+\dkn)R^6 x^2 \tau^4-2 \Dkn R^3 x
\tau^2\{\arbh
\dkn[R^3(\arbh x+Ry)+ \arbh z^2(-Rx+\arbh y)] \nonumber \\
&&-R^3x\tau\}+\Dkn^2\{\arbh^2 \dkn[R^3(\arbh x+Ry)+ \arbh z^2(-Rx+\arbh y)]^2 \nonumber \\
&&+(R^8-R^6x^2+2\arbh^2R^4z^2+ \arbh^4z^4)\tau^2\}),
\end{eqnarray}
\begin{eqnarray}
g_{12}&=&\sigma^2 (\frac{R^6xy(\Dkn-\tau)^2}{\tau^2}+\dkn\{\arbh
\Dkn[R^3(\arbh x+Ry)+ \arbh (-Rx+ \arbh y)z^2] \nonumber \\
&&-R^3x\tau^2\}\{ \arbh \Dkn[R^3(Rx- \arbh y)+ \arbh ( \arbh
x+Ry)z^2]+R^3y\tau^2\}),
\end{eqnarray}
\begin{eqnarray}
g_{13}&=&\frac{z \sigma^2
\tau}{R}(R^5x(\Dkn-\tau)^2\tau^2-\dkn[\arbh^2\Dkn(-R^2+z^2)+R^2
\tau^2]\{-\arbh \Dkn[R^3( \arbh x+Ry) \nonumber \\
&& + \arbh (-Rx+ \arbh y)z^2]+R^3x\tau^2\}),
\end{eqnarray}
\begin{eqnarray}
g_{22}&=&-\sigma^2(\arbh^2\dkn
\Dkn^2[R^3(Rx- \arbh y)+ \arbh ( \arbh x +Ry)z^2]^2+\Dkn\{R^6[\Dkn(R-y)(R+y) \nonumber \\
&&+2 \arbh \dkn y(Rx- \arbh y)]+2 \arbh^2R^3[\Dkn R+\dkn y( \arbh
x+Ry)]z^2\} \arbh^4 \Dkn z^4 )
\tau^2 \nonumber \\
&&+2 \Dkn R^6 y^2 \tau^3+(-1+\dkn)R^6y^2\tau^4,
\end{eqnarray}
\begin{eqnarray}
g_{23}&=&\frac{z \sigma^2 \tau}{R}(R^5
y(\Dkn-\tau)^2\tau^2-\dkn[\arbh^2\Dkn(-R^2+z^2)+R^2\tau^2] \nonumber \\
&& \times \{\arbh \Dkn[R^3(Rx- \arbh y)+ \arbh ( \arbh
x+Ry)z^2]+R^3y\tau^2\}),
\end{eqnarray}
\begin{eqnarray}
g_{33}&=&\frac{-1}{R^2}(\sigma^2
\tau^2\{\Dkn^2(R-z)(R+z)[R^8+\arbh^4(-1+\dkn)R^2z^2-\arbh^4\dkn z^4]
\nonumber \\
&&+2\Dkn R^2 z^2 [R^4+\arbh^2(R^2-\dkn R^2+\dkn
z^2)]\tau^2+(-1+\dkn)R^4z^2\tau^4\}),
\end{eqnarray}
and $g_{\alpha \beta}=g_{\beta \alpha}$. Also,
\begin{equation}
\dkn=\frac{R^2 \left(2 \mrbh R-q^2\right)}{R^4+\arbh^2 z^2},
\end{equation}
\begin{equation}
\tau=\arbh^2+R^2,
\end{equation}
\begin{equation}
\sigma=\frac{1}{\Dkn \tau (R^4+\arbh^2z^2)}.
\end{equation}

\vspace{5mm} \noindent \Large{{\bf Appendix 2}} \normalsize
\vspace{5mm}

The components of the Kerr--Sen metric obtained from the line
element \r{KS_le_KS} expressed in Kerr--Schild Cartesian form are
\begin{eqnarray}
g_{00}=1-\dks,
\end{eqnarray}
\begin{eqnarray}
g_{01}&=&\Phi\{(-1+\dks)S^3x[-2\arbh^2\Dks S+\nu(-\Dks
\beta+\rho\nu)]+2\Dks \St[-\arbh \dks y(S^4+\arbh^2z^2) \nonumber \\
&&+x(S^4+\arbh^2\dks z^2)\St]\},
\end{eqnarray}
\begin{eqnarray}
g_{02}&=&\Phi\{(-1+\dks)S^3y[-2\arbh^2\Dks S+\nu(-\Dks
\beta+\rho\nu)]+2\Dks \St[\arbh \dks x(S^4+\arbh^2z^2)\nonumber \\
&&+y(S^4+\arbh^2\dks z^2)\St]\},
\end{eqnarray}
\begin{eqnarray}
g_{03}&=&\nu \Phi[\Dks z(2\{S^4+\arbh^2[-(-1+\dks)S^2+\dks
z^2]\}-(-1+\dks)S(\arbh^2+S^2)\beta)\nonumber \\
&&+(-1+\dks)S(\arbh^2+S^2)z\rho\nu],
\end{eqnarray}
\begin{eqnarray}
g_{11}&=&\Phi^2\{S^6x^2[-(-1+\dks)\{2\arbh^2\Dks S+\nu[\Dks
\beta-\rho\nu]\}^2+\Dks^2(2\arbh^2S+\beta\nu)^2\zeta]\nonumber \\
&&+4\Dks\St[\arbh\dks S^3xy(S^4+\arbh^2z^2)[-2\arbh^2\Dks
S+\nu(-\Dks\beta+2S\nu+\beta\nu)]\nonumber \\
&&-[\arbh^8\Dks z^4+\arbh^6\Dks z^2(2S^4+\dks y^2z^2)+\arbh^4\Dks
S^4[S^4+2\dks(-x^2+y^2)z^2]\nonumber \\
&&+S^7x^2\nu[\rho\nu-\Dks\beta(1+\zeta)]+\arbh^2S^3(\dks
x^2z^2\rho\nu^2-\Dks\{\dks x^2z^2\beta\nu\nonumber \\
&&+S^5[-\dks y^2+2x^2(1+\zeta)]\})]\St+2\arbh^3\dks\Dks
xyz^2(S^4+\arbh^2z^2)\St^2-\Dks[2\arbh^2S^8 \nonumber \\
&&+2\arbh^6z^4+\arbh^4(4S^4z^2+\dks
x^2z^4)-S^8x^2(1+\zeta)]\St^3-\Dks(S^4+\arbh^2z^2)^2\St^5]\},
\end{eqnarray}
\begin{eqnarray}
g_{12}&=&\Dks^2\nu^2\Phi^2[\frac{S^6xy\rho^2(\Dks-\nu)^2}{\Dks^2}+S^6xy\rho^2\zeta+
\frac{1}{\Dks^2\nu^2}(\dks(S^3x\{2\arbh^2\Dks
S\nonumber \\
&&+\nu[\Dks\beta-\rho\nu]\}+2\arbh\Dks\St(y(S^4+\arbh^2z^2)- \arbh
xz^2\St))\{S^3y\{-2 \arbh^2\Dks
S\nonumber \\
&&+\nu[-\Dks\beta+\rho\nu]\}+2\arbh\Dks\St[x(S^4+\arbh^2z^2)+ \arbh
yz^2\St]\})],
\end{eqnarray}
\begin{eqnarray}
g_{13}&=&\Dks^2z\nu^2\Phi^2\{\frac{S^4(\arbh^2+S^2)x\rho^2(\Dks-\nu)^2}{\Dks^2}+S^4(\arbh^2+S^2)x\rho^2\zeta\nonumber \\
&&-\frac{1}{\Dks^2\nu} [\dks\{\Dks
S^3\beta+\arbh^2\Dks[-2z^2+S\rho]-S(\arbh^2+S^2)\rho\nu\}(S^3x\{2\arbh^2\Dks
S\nonumber \\
&&+\nu[\Dks\beta-\rho\nu]\}+2\arbh\Dks\St[y(S^4+\arbh^2z^2)- \arbh x
z^2\St])]\},
\end{eqnarray}
\begin{eqnarray}
g_{22}&=&-\Phi^2\{S^6y^2\{-\Dks^2(2\arbh^2S+\beta\nu)^2\zeta+(-1+\dks)[2\arbh^2\Dks
S+\nu(\Dks\beta-\nu\rho)]^2\}\nonumber \\
&&+4\Dks\St[\arbh\dks S^3xy(S^4+\arbh^2z^2)[-2\arbh^2\Dks
S+\nu(-\Dks\beta+\nu\rho)]+\St(\arbh^8\Dks z^4\nonumber \\
&&+\arbh^6\Dks z^2(2S^4+\dks x^2z^2)+\arbh^4\Dks
S^4[S^4+2\dks(x-y)(x+y)z^2]\nonumber \\
&&+S^7y^2\nu[-\Dks\beta(1+\zeta)+\nu\rho]+\arbh^2S^3[-2\Dks
S^5y^2(1+\zeta)+\dks(\Dks S^5x^2\nonumber \\
&&-\Dks y^2z^2\beta\nu+y^2z^2\nu^2\rho)]+\Dks \St\{2\arbh^3\dks
xyz^2(S^4+\arbh^2z^2)+[2\arbh^2S^8+2\arbh^6z^4\nonumber \\
&&+\arbh^4(4S^4z^2+\dks
y^2z^4)-S^8y^2(1+\zeta)]\St+(S^4+\arbh^2z^2)^2\St^3\})]\},
\end{eqnarray}
\begin{eqnarray}
g_{23}&=&\Dks^2 z \nu^2\Phi^2[\frac{S^4(\arbh^2+S^2)y(\Dks-\nu)^2
\rho^2}{\Dks^2}+S^4(\arbh^2+S^2)y\zeta\rho^2\nonumber \\
&&+\frac{1}{\Dks^2\nu}(\dks\{\arbh^2[-2\Dks
z^2+S(\Dks-\nu)\rho]\nonumber \\
&&+S^3(\Dks\beta-\nu\rho)\}\{S^3y(-2\arbh^2\Dks
S+\nu(-\Dks\beta+\nu\rho))\nonumber \\
&&+2\arbh\Dks \St[x(S^4+\arbh^2z^2)+ \arbh yz^2\St]\})],
\end{eqnarray}
\begin{eqnarray}
g_{33}&=&\nu^2\Phi^2\{z^2(\Dks^2
S^6[4S^2(1+\zeta)+4S\beta(1+\zeta)+\beta^2(1+\zeta-\dks)]\nonumber \\
&&-2\arbh^6S[2(-1+\dks)S^2(-\Dks+S^2)+2\dks \Dks
z^2\nonumber \\
&&+(-1+\dks)S(-\Dks+S^2)\beta]\rho-\arbh^8(-1+\dks)S^2\rho^2+2\arbh^2\Dks
S^3\{\Dks[2\dks
z^2\beta\nonumber \\
&&-2S^2\beta(-2+\dks-2\zeta)+4S^3(1+\zeta)+S\beta^2(1-\dks+\zeta)]+S^3(\dks\beta-\rho)\rho\}\nonumber \\
&&+\arbh^4\{\Dks^2(-4\dks z^4+4\dks S
z^2\beta+4S^3\beta(1-\dks+\zeta)\nonumber \\
&&+S^4(4-4\dks+4\zeta)+S^2(8\dks
z^2+\beta^2-\dks\beta^2+\beta^2\zeta))+4\Dks S^3[(-2+\dks)S^2\nonumber \\
&&-\dks
z^2+(-1+\dks)S\beta]\rho-(-1+\dks)S^6\rho^2\})-2[2\Dks^2(S^4+\arbh^2z^2)^2\nonumber \\
&&+\Dks S(\arbh^2+S^2)z^2(2\{S^4+\arbh^2[-(-1+\dks)S^2+\dks
z^2]\}-(-1+\dks)S(\arbh^2\nonumber \\
&&+S^2)\beta)\rho+\arbh^2(-1+\dks)(\arbh^2S+S^3)^2z^2\rho^2]\St^2\nonumber \\
&&-(-1+\dks)S^2(\arbh^2+S^2)^2z^2\rho^2\St^4\},
\end{eqnarray}
and $g_{\alpha \beta}=g_{\beta \alpha}$. Also,
\begin{equation}
\dks=\frac{2 \mrbh \St^2 S}{\St^4+\arbh^2z^2},
\end{equation}
\begin{equation}
\nu=\arbh^2+\St^2,
\end{equation}
\begin{equation}
\rho=2S+\beta,
\end{equation}
\begin{equation}
\zeta=\frac{\beta^2(\St^2+\arbh^2 \cos^2(\theta))}{(\beta^2+4
\St^2)\Dks},
\end{equation}
\begin{equation}
\Phi=\frac{1}{2 \Dks (S^4+a^2 z^2) \nu \St}\,.
\end{equation}


\newpage


\newpage

\end{document}